\documentclass{article}%
\usepackage{amsmath}
\usepackage{amsfonts}
\usepackage{amssymb}
\usepackage{graphicx}%
\begin{document}

\title{Localization-Entropy from Holography on Null-Surfaces and the Split Property}
\author{Bert Schroer\\CBPF, Rua Dr. Xavier Sigaud 150 \\22290-180 Rio de Janeiro, Brazil\\and Institut fuer Theoretische Physik der FU Berlin, Germany}
\date{December 2007}
\maketitle

\begin{abstract}
Using the conformal equivalence of translational KMS states on chiral theories
with dilational KMS states obtained from restricting the vacuum state to an
interval (the chiral inversion of the Unruh-effect) it was shown in a previous
publications that the diverging volume (length) factor of the thermodynamic
limit corresponds to the logarithmic depencence on a decreasing attenuation
length associated with the localization-caused vacuum polarization cloud near
the causal boundary of the localization region. Far from being a coincidence
this is a structural consequence of the fact that both operator algebras, the
global thermal and the locally restricted ground state algebra are of the same
unique von Neumann type ( monad") which is completely different from that met
in Born-localized quantum mechanical algebras. Together with the technique of
holographic projection this leads to the universal area proportionality.

The main aim in this paper is to describe a derivation which is more in the
spirit of recent work on entanglement entropy in condensed matter physics,
especially to that of the replica trick as used by Cardy and collaborators.
The essential new ingredient is the use of the split property which already
has shown its constructive power in securing the existence of models of
factorizing theories.

\end{abstract}

\section{Review of known facts}

Whereas quantum mechanics (QM) has reached its conceptual maturity a long time
ago, relativistic quantum field theory (QFT) is a sophisticated and
computational very demanding theory whose conceptual closure remains still a
future project. This is evident from the scarcity of mathematically reliable
results about interacting models. After almost four decades of stalemate,
there has been some modest progress on the central ontological problem of QFT,
namely the existence and nonperturbative constructions of interacting models
with noncanonical short distance (strictly renormalizable) behavior.

In a classic paper \cite{Ha-Sw} of the 60s entitled "When does a Quantum Field
Theory describe Particles?", the issue of the phase space degrees of freedom,
which finally led to the split property, entered QFT for the first time. In
that paper it was shown that, contrary to naive analogies claiming that QFT is
essentially relativistic QM, the phase space structure of QFT is sufficiently
different from QM as to merit attention for a better understanding of the
field-particle relation\footnote{Interacting fields do not create one-particle
states; in fact they do not even create states which contain only a finite
number of particles.} \cite{legacy}. As the title reveals, the authors
conjectured that this property is indispensable for the understanding of
asymptotic completeness of particle states i.e. of the equality of the
Wigner-Fock space with the Hilbert space defined by the quantum fields. This
structure of the Hilbert space and its connection to the local properties of
the observables via scattering theory were expected to play an important role
in establishing the existence of nontrivial models. 

The phase space properties of that paper were later significantly sharpened
\cite{Bu-Wi} and used in order to prove that a theory in the ground state with
a suitable phase space structure also exists in a thermal equilibrium state
\cite{Bu-Ju}, in other words one can directly pass from the theory in its
ground state representation to the thermal setting without running through a
different "Thermo-field" quantization procedure. The nontrivial aspect of the
argument lies in overcoming the barrier of the inequivalence of the two
representations which results in a difference of the von Neumann type of the
operator algebras. This is where the \textit{split property} as a consequence
of the phase space requirement plays a crucial role.

Recently it was established that the issue of the asymptotic particle
completeness and the mathematical existence of QFTs are indeed inexorably
linked; within the class of factorizing models they are consequences of
verifiable phase space properties. Hence within this setting, the Haag-Swieca
conjecture about phase space properties leading to the complete particle
interpretation within a mathematically controlled QFTs was vindicated. \ Since
the conceptual and mathematical tools are very different from the better known
Lagrangian quantization and functional integral setting, some additional
introductory remarks on this topic may be helpful. \ 

Factorizing models are two-dimensional models whose S matrix is purely
elastic, a property which is very atypical in QFT. The only known way in which
thiis can be consistent with the multi-particle matrix elements of $S$
factorize in an appropriate way (consistent with the cluster property and
macro-causality) in terms of $S^{(2)}.$ Elasticity is a property which is
known to contradict the structure of interacting QFTs in higher spacetime
dimensions \cite{Aks}. In this factorizing context the requirements of the old
(aborted) S-matrix bootstrap approach really work in the sense that it
possible to classify unitary, Poincar\'{e} invariant $S^{(2)}$ with the very
nontrivial crossing property \cite{Ka1}. 

But very different from the dreams the protagonists of the bootstrap had in
the 60s, these principles are not only incapable to select a unique theory of
everything (TOE), but they rather act in the opposite direction of generating
more models than can be "baptized"  by local interaction Lagrangians. The
infinitely many bootstrap S-matrices can be arranged into families according
to symmetry principles and they possess uniquely associated QFTs whose
formfactors were constructed within the bootstrap-formfactor program.\ An
important technical tool was added in form of the Zamolodchikov-Faddeev
algebra which represents a specific nonlocal modification of on-shell
creation/annihilation operators.

The absence of on-shell particle creation (through scattering) is reminiscent
of particle number-conservation in QM. But the characteristic property of
interactions in QFT is the presence of vacuum polarization clouds in compactly
localized states and less the on-shell reation\footnote{The relativistic QM of
\textit{direct interactions} allows the introduction of creation channels "by
hand", whereas it is not possible to add vacuum polarization, i.e. there is no
passage from the relativistic direct interaction setting to QFT.}.

The generators of the Z-F algebra obtained a physically important spacetime
when it was recognized that they play a role in the covariant but nonlocal
generating fields of the wedge-localized operator algebras \cite{AOP}. With
this spacetime interpretation it became possible to reformulate the
bootstrap-formfactor program in terms of local properties based ob the concept
of PFGs. This notion of vacuum-\textbf{p}olarization-\textbf{f}ree-\textbf{g}%
enerators, i.e. of operators which generate vacuum-polarization-free one
particle states if acting on the vacuum, turned out to be extremely useful as
a local indicator for the absence/presence of interactions. If a subwedge
localized PFG exists in a theory, then it possesses a description in terms of
a free field.

PFGs always exist as a consequence of scattering theory within the global
algebra which lives on full Minkowski spacetime. Modular operator theory
assures the existence of wedge-localized PFGs i.e. the wedge region is the
best compromise between fields (as carriers of localization) and particles. In
general these wedge generators have very bad domain properties which prevent
their successive application to the vacuum. To arrive at a manageable
situation one demands temperateness of the wedge-localized PFGs
\cite{Bo-Bu-Sc} and shows that they lead to pure elastic S-matrix of
two-dimensional theories. With a mild additional assumption one then arrives
at factorizing theories. In this way one gets a very different spacetime
characterization of field theoretic integrability in which vacuum polarization
clouds play the central role. The advantage of this construction is that one
always maintains the connection with the general physical principles which
underlie all QFTs and, last not least, for the first time one obtains an
existence proof for noncanonical QFTs \cite{Lech}.

In this context it should be recalled that standard methods of showing the
existence of pointlike covariant off-shell field generators in terms of
on-shell Z-F operators remained inconclusive since formfactors are at best
\textit{bilinear forms of would-be operators}; to obtain the existence of bona
fide operators and their algebras one has to use alternative methods which
avoid the use of pointlike field coordinatization and address the
nontriviality of the local algebras in a more intrinsic manner. The experience
with divergent perturbative series casts doubts about the convergence status
of the two known formally exact nonperturbative series representation of
interacting quantum fields namely the Glaser-Lehmann-Symanzik (GLZ) series in
terms of incoming free fields and the closely related series\footnote{The
convergence of these series within the Wigner-Fock space of the asumptotic
particles would justify the interpretation of formfactors as multi-particle
matrix elements of operators in the W-F Hilbert space.} for fields in
factorizing models in terms Zamolodchikov-Faddeev (Z-F) creation/annihilation
operators of representation of Heisenberg fields. Hence the algebraic
existence proof is much more than an elegant reformulation of an already
existing result. We will return to these ontological issues of interacting
QFTs within a more detailed setting in the third section.

The algebraic existence proof for two-dimensional factorizing QFTs uses the
simplicity of temperate PFGs as generators of wedge-algebra in an essential
way. A perturbative construction of wedge generators in higher dimensions,
where no temperate wedge-localized PFGs are available, is an interesting open
problem. But it would almost certainly re-introduce those unsolved convergence
problems connected with the perturbative series. Its main advantage is
expected to be the enlargement the realm of renormalizable interactions
(finite-parametric QFTs) through avoidance of pointlike fields. In \cite{MSY}
it was shown that already the use of covariant stringlike localized free
fields instead of pointlike fields does improve the short distance behavior so
that many more interactions acquire the formal renormalizabilty status in
terms of power counting. One expects that starting with wedge-like localized
generators one reaches the frontier of renormalizability i.e. the most general
class of interactions for which perturbation theory remains polynomially
bounded and permits a finite parametric description which is stable under the
action of some suitable defined renormalization group.

A promising rigorous \textit{nonperturbative} idea in higher dimensions
consists in working with \textit{holographic mappings of wedge-localized
algebras}. Holographic projections in which a localized bulk algebra is
projected onto the null horizon of the causally completed bulk lead to
significant simplifications in the description of localization-caused
properties as energy and entropy densities. Naturally simplifications which do
not modify the content of models can only consist in looking at a model from a
different vantage point, so that certain aspects one wants to focus on become
simpler at the expense of complicating other aspects. QFT still hides many
surprises but there are no miracles.

The most prominent alternative approach with this aim is quite old and known
under the name of "lightcone quantization" (closely connected to the
"p$\rightarrow\infty$ frame" method). The name suggests that it was originally
thought of as an alternative quantization to the standard Lagrangian approach
from which one expected a simpler understanding of certain high-energy
aspects. Since a change of quantization generally leads to a change of the QFT
model, this raises the question if, and in what way, the new method is
conceptually as well as computationally related to the standard formulation
\cite{Leut}\cite{Drie1}, a problem which unfortunataly was not addressed in
most papers. \ 

Lightcone quantization or (in the more appropriate terminology used in this
paper) \textit{lightfront restriction} suffers the same formal limitations as
equal time commutation relations, namely it becomes meaningless in the
Wightman representation using x-space correlation functions (Wightman
functions) whenever the wave function renormalization constant $Z$ ceases to
be finite and non-vanishing.

The fact that in these cases of infinite wave function renormalization the
canonical equal time formalism breaks down has no serious consequences since
physical requirements in QFT do not demand that the commutation relations of
pointlike fields ought to allow an equal time restriction. Within the
perturbative setting this is fully taken into account in the so called
\textit{causal perturbation theory} which is only subject to the more general
and much weaker requirement of \textit{renormalizability.}

In the following section we use the simplicity of lightfront restriction for
free fields in order to explain some important concepts which were absent in
the old lightcone quantization work and which have no natural formulation in
Lagrangian quantization. In the same section some algebraic concepts, which
highlight the completely different nature of localized algebras in QFT from
those in QM, will be introduced. The distinction between quantum mechanical
entanglement and the thermal aspects of localized algebras in QFT (which being
monads have no pure states \cite{interface} at all), whose causal horizon is
surrounded by a vacuum polarization cloud with an attentuation length
$\varepsilon$ depending on the variable splitting arrangement, are important
manifestations of this significant conceptual difference between QM and QFT.

The third section presents the algebraic holography in the presence of
interactions. In that case the connection between the \textit{algebraic
holography,} which uses the notions of \textit{modular inclusions} and
\textit{modular intersections} of wedge algebras, and the (at the present time
less rigorous) \textit{projective holography }of pointlike fields (which
requires a mass-shell representation of the interacting fields in terms of an
infinite series in the incoming or the Z-F creation/annihilation operators
with an unclear convergence status) still needs further clarification.

Some observations, which suggest the way in which the two holographies are
related, can be made in the setting of factorizing models and will be
presented in the fourth section. There we also comment on the conceptual
difference between the holographic projection and the critical limit theory of
the bulk (the representative of the universality class ) which is known to be
a conformal QFT in a different Hilbert space. This has an interesting but
still somewhat speculative relation to the issue of Zamolodchikov's
deformation of chiral models into factorizing theories.

In the fifth section we remind the reader of the close structural algebraic
relation of the thermodynamic limit of a heat bath system and the thermal
aspects of localization in QFT. We show that for chiral algebras the length (=
one-dimensional volume) factor passes via conformal covariance to the
logarithmic attenuation factor. The transversely extended chiral theory, which
results from holographic projection of the bulk, turns out to be the
\textit{raison d'etre for the universal area proportionality of
localization-entropy}.

In order to arrive at a completely intrinsic alternative derivation via direct
use of the split property, we adapt the replica idea of the derivation of
localization-entropy as used by the condensed matter physicists \cite{Ca} to
the present setting. The result is a conceptually transparent implementation
of the replica trick which avoids the very artful but nevertheless metaphoric
functional integral representations which will be presented in the sixth section.

The concluding section contains some speculative remarks about the future of
QFT which are of a more philosophical nature

\section{Holographic projection for free fields, an illustrative example}

The shortcomings of the old "lightcone quantization" become more explicit if
one compares it with its modern successor which, for reasons which become
obvious, will be referred to as "lightfront holography" (LFH). The main idea
of LFH can be illustrated with the help of the mass shell representation of
the free scalar field $A(x)$ and its restriction to the wedge region
\cite{CQG1}\cite{CQG2}
\begin{align}
A(x) &  =\frac{1}{\left(  2\pi\right)  ^{\frac{3}{2}}}%
%TCIMACRO{\dint }%
%BeginExpansion
{\displaystyle\int}
%EndExpansion
(e^{ipx}a^{\ast}(p)+h.c.)\frac{d^{3}p}{2p_{0}}\label{free}\\
A_{W}(r,\chi;x_{\perp}) &  \equiv A(x)|_{W}=\frac{1}{\left(  2\pi\right)
^{\frac{3}{2}}}%
%TCIMACRO{\dint }%
%BeginExpansion
{\displaystyle\int}
%EndExpansion
(e^{im_{eff}rch(\chi-\theta)+ip_{\perp}x_{\perp}}a^{\ast}(\theta,p_{\perp
})+h.c.)\frac{d\theta}{2}dp_{\perp}~\nonumber\\
m_{eff} &  =\sqrt{m^{2}+p_{\perp}^{2}}%
\end{align}
where we have chosen as our standard wedge the $x_{1}-x_{0}$ wedge (invariant
under the $x_{1}-x_{0}$ boost subgroup) and parametrized the longitudinal
coordinates in terms of x-space and p-space rapidities. The encoding of
structural properties of the $\mathcal{A}(W)$ bulk into simpler properties of
its holographic projection onto the causal horizon $\mathcal{A}(H(W))$ can
then be described by passing to the limit\footnote{Whereas the holographic
projection of fields in the mass-shell representation (see below) could have
been done directly by setting $x_{-}=0,$ algebraic holography needs the
setting of modular operator theory which only works for localized algebras
which fulfill the Reeh-Schlieder property.} $r\rightarrow0,$ $\chi
\rightarrow\infty$ such that $x_{-}=0,$ $x_{+}>0$ and finite. In this limit
the mass looses its physical significance and becomes a mere placeholder for
keeping track of the engineering dimension
\begin{align}
&  A_{H(W)}(x_{+},x_{\perp})=\frac{1}{\left(  2\pi\right)  ^{\frac{3}{2}}}%
%TCIMACRO{\dint }%
%BeginExpansion
{\displaystyle\int}
%EndExpansion
(e^{im_{eff}x_{+}e^{\theta}+ip_{\perp}x_{\perp}}a^{\ast}(\theta,p_{\perp
})+h.c.)\frac{d\theta}{2}dp_{\perp}\label{hol}\\
&  \left\langle \partial_{x_{+}}A_{H(W)}(x_{+},x_{\perp})\partial_{x\prime
_{+}}A_{H(W)}(x_{+}^{\prime},x_{\perp}^{\prime})\right\rangle \simeq\frac
{1}{\left(  x_{+}-x_{+}^{\prime}+i\varepsilon\right)  ^{2}}\cdot
\delta(x_{\perp}-x_{\perp}^{\prime})\nonumber\\
&  \left[  \partial_{x_{+}}A_{H(W)}(x_{+},x_{\perp}),\partial_{x\prime_{+}%
}A_{H(W)}(x_{+}^{\prime},x_{\perp}^{\prime})\right]  \simeq\delta^{\prime
}(x_{+}-x_{+}^{\prime})\cdot\delta(x_{\perp}-x_{\perp}^{\prime})\nonumber
\end{align}
For convenience we have taken the lightray derivatives of the generating
fields; this saves us the usual ritual of restricted testfunction spaces for
zero mass chiral free fields; upon \textit{Haag-dualization} the algebras
generated by the derivative fields are identical to those defined with
modified testfunction smearing.

The lightfront fields $A_{LF}(x_{+},x_{\perp})$ are obtained by linearly
extending the $A_{H(W)}(x_{+},x_{\perp})$ to all values of $x_{+}.$ The fields
$A_{W}(r,\chi;x_{\perp})$ and $A_{H(W)}(x_{+},x_{\perp})$ (respectively their
derivatives) generate operator algebras\footnote{The relation between free
field generators and operator algebras is defined in terms of the Weyl
generators but these well-known technical points are left out in these notes.}
$\mathcal{A}(W)$ and $\mathcal{A}(H(W)).$ The local algebras generated by the
derivatives are smaller, but the extension via Haag dualization restores
equality \cite{GLW}. It is fairly easy to see that%
\begin{equation}
\mathcal{A}(W)=\mathcal{A}(H(W))\label{h}%
\end{equation}
In fact by using the relation between the on-shell restriction of W-supported
smearing functions and smearing on $H(W),$ the generators can be directly
placed into correspondence.

This identity is surprising at first sight since $\mathcal{A}(H(W))$ or its
linear extension $\mathcal{A}(LF)$ does not distinguish between a massive and
a massless theory. However this identity does not extend to compactly
localized subalgebras; the knowledge of lightfront generators (\ref{hol}) on
one lightfront only does not suffice to reconstruct compactly localized
algebras in the bulk. Vice versa on cannot construct the local substructure on
the horizon from the bulk substructure  of its associated wedge. With
additional information about certain LF changing action of the Poincar\'{e}
group, and by taking algebraic intersections and unions, one can however
recover the local bulk structure and its pointlike generating field
coordinatizations (which includes besides the free fields also its
Wick-monomials). The reason behind these exact connections is that, different
from the critical limit of a massive theory which leads to a conformal
invariant massless theory, the Hilbert space and certain noncompact localized
subalgebras are shared between the bulk and the lightfront, a fact which in
our example is obvious since we never changed the Wigner particle
creation/annihilation operators for our massive particles i.e. the full
content of the representation of the Poincar\'{e} group remained encoded in
both descriptions.

The main point of lightfront holography (as we will denote the present setting
in order to distinguish it from the old lightcone quantization) is precisely a
radical change of spacetime ordering while maintaining the material substrate.
Since the spacetime ordering of matter is crucial for its physical
interpretation, certain concepts for which the bulk description was important,
as e.g. scattering theory of particles, become blurred in the holographic
projection and concepts like localization entropy/energy, which are essential
for the understanding of the area behavior of entropy on causal/event
horizons, are not very accessible from a pure bulk point of view. Already the
above free field calculation reveals that the holographic generator $A_{H(W)}$
is in many aspects simpler since it generates a transversely extended chiral
theory \ with no transverse vacuum fluctuation. It can be used to generate
$\mathcal{A}(W),$ but is is of no use for generating all the subalgebras
contained in $\mathcal{A}(W),$ for this one would have to use the bulk free
field $A_{W}(x)$.

There is no direct reconstruction of the bulk field $A_{W}$ from the boundary
field $A_{H(W)},$ rather one has to pass through intermediate purely algebraic
steps as intersecting wedge algebras to obtain double cone algebras
$\mathcal{A(O})$ which for arbitrary small $\mathcal{O}^{\prime}s$ in turn
lead back to the pointlike covariant fields which generate all subalgebras.
Hence a constructive approach in this setting starts from the algebraic
structure of a wedge algebra $\mathcal{A}(W)$ in general position (obtained
from a special one by applying the Poincar\'{e} group) and the aim is to
obtain the compactly localized double cone algebras $\mathcal{A}(C$) from
algebraic intersection. If these intersection algebras are trivial (scalar
multiples of the identity) then there are no local observables we say that the
model does not exist as local QFT.

Opposite to the standard approach which moves from pointlike fields to more
extended observables, the direction of the algebraic approach is
outward$\rightarrow$inward; the local field only appears at the end in its
role as the generator of algebras for all spacetime regions. These algebraic
constructions have been backed up by explicit calculations which established
the existence \cite{Lech} of a two-dimensional class of so-called factorizing
models whose S-matrix and formfactors had already been known before.

The rather direct relation between free fields and their holographic
generators holds only for linear nullsurfaces i.e. the lightfront. The causal
horizon of a double cone $C$ (the intersection of two lightcones) has also
chiral generators in lightray direction and vacuum-polarization-free
transverse angular generators, but they cannot be obtained by a restriction
procedure on free fields which generalizes the above limit leading to
$A_{H(C)}(x_{H(C)})$. For conformally invariant models they can be obtained by
applying the relevant conformal transformation (i.e. that one which maps the
wedge into a double cone) to the $\mathcal{A}_{H(W)}$ generator \cite{CQG2}.
Sine the holographic projections are conformally invariant even if the bulk
theory is not conformal, the application of conformal maps between different
null-horizons even in case where the bulk is not conformal has a certain
plausibility, but a rigorous mathematical justification is still missing.

The structure of the localized operator algebras in QFT are very different
from those one meets in QM. Whereas the operator algebra representing the
total (generally infinitely extended i.e. open) system\footnote{In order to
have a common setting for QM and QFT one should use the Wigner-Fock
multiparticle formulation of QM.} at zero temperature in both cases is the
irreducible algebra of all bounded operators in a Hilbert space $B(H),$ this
does not extend to local subalgebras of QFT. Using the notation $\mathcal{N}%
(V)\subset B(H)$ for the von Neumann subalgebra associated to the 3-dim.
region $V\subset\mathbb{R}^{3}$ and $V^{\prime}=V\backslash\mathbb{R}^{3}$ its
complement, the quantum mechanical tensor factorization reads%
\begin{align*}
\mathcal{N}(\mathbb{R}^{3}) &  =\mathcal{N}(V)\otimes\mathcal{N}(V^{\prime
}),~H=H_{V}\otimes H_{V}\\
\mathcal{N}(\mathbb{R}^{3}) &  =B(H),~\mathcal{N}(V)=B(H_{V}),~\mathcal{N}%
(V^{\prime})=B(H_{V})
\end{align*}
Pure states of the form $\sum\lambda_{i}\left\vert \psi_{i}\right\rangle
\left\langle \psi_{i}^{\prime}\right\vert $ which are superpositions with
respect to the tensor factorization are called entangled (with respect to the
tensor decomposition) and the averaging over the unobserved degrees of freedom
in the say observational inaccessible region $V^{\prime}$ leads to a density
matrix $\rho_{V}$ for the observer confined to $V.$ The impurity of the state
can be measured in terms of the von Neumann entropy associated to $\rho_{V}$
but there is no thermal manifestation of Born localization i.e. the
uncertainty relation is not related to thermal behavior. \textit{The quantum
mechanical entropy resulting from "Born"-entangled states on a tensor product
lead to a "cold" entropy in the sense of informations theory \cite{interface}.
The localization entawking radiation. }As there is no problem of loss of
information in properly understood black hole physics, the entanglement in
tensor products of QM cannot be used to generate thermal manifestations.

The situation changes radically if one passes from QM to QFT \cite{Yng}%
\cite{interface} In that case a covariant localized subalgebra $\mathcal{A(O)}%
\subset\mathcal{A}(M),$ $\mathcal{O}\subset M=~$Minkowski spacetime. In this
case the complement is replaced by the causal disjoint $\mathcal{O}^{\prime}$
whose algebraic counterpart is the commutant i.e. $\mathcal{A(O}^{\prime
})=\mathcal{A(O})^{\prime}$ (Haag duality, a property which can be always
achieved by suitably enlarging the local algebras) and the generating and
causal commuting property%
\begin{align}
\mathcal{A(O)\vee A(O}^{\prime})  &  =\mathcal{A}(M)=B(H)\\
\left[  \mathcal{A(O)},\mathcal{A(O}^{\prime})\right]   &  =0\nonumber
\end{align}
In the classification of Murray and von Neumann field theoretic algebras as
wedge-localized algebras $\mathcal{A(}W\mathcal{)}$ and its causal complement
$\mathcal{A}(W^{\prime})=\mathcal{A}(W)^{\prime}$ are still factor
algebras\footnote{Under very mild assumptions the algebras for subwedge
localization regions $\mathcal{A}(\mathcal{O})$ are of the same type as
$\mathcal{A}(W).$}, but there is no tensor factorization and hence no
prerequisite for defining the analog of the above entanglement.

In fact the local factor algebra does not admit any pure state, rather
\textit{all states are impure}. Any attempt to go ahead and ignore this
difference will lead to infinities and ill-defined expressions for measuring
the impurity in terms of an entropy. Although we will not enter mathematical
subtleties, the basic reason for this unusual state of affairs can be traced
back to the radically different nature of these local covariant subalgebras of
which the most prominent example in this article is $\mathcal{A}(W).$ They are
all (as long as the localization region has a nontrivial causal complement)
isomorphic to the unique hyperfinite type III$_{1}$ von Neumann factor which
for reasons which will become later will be shortly referred to as the
\textit{monad} in the rest of this article. It is however not our aim to enter
a systematic mathematical discussion about operator algebras \cite{Su}%
\cite{Yng}, rather we will pay attention to those aspects of operator algebras
which are important for the problematization of holography and
thermal/entropic aspects of localization \cite{CQG1}\cite{CQG2}.

Although, as mentioned before, monads does not arise in global zero
temperature algebras $\mathcal{N}(\mathbb{R}^{3})$ or $\mathcal{A}(M),$ they
do however make their appearance in the thermodynamic limit of finite
temperature systems. This preempts the thermal aspects of localization in QFT
on the very fundamental level of single operator algebras. The vastly
different behavior of localized subalgebras in QM and QFT is of course a
reflection of the difference between "Born localization" \cite{interface} and
modular localization \cite{MSY}. The origin of the terminology "modular" will
become clearer in the sequel. but in the present context it stands for an
intrinsic formulation (i.e. independent of the myriads of field
coordinatizations which a particular model admits). "Born localization" is the
standard quantum mechanical localization which is directly associated with
projection operators and probabilities to find the system at a given time in a
certain spatial region. In the relativistic context the Born localization is
usually referred to as the Newton-Wigner localization. Its lack of local
covariance makes its rather useless for the problems of causal propagation
over finite spacetime distances.

But instead of pointing to its shortcomings outside QM it is more important in
the context of our present discussion to stress the fact that it is
asymptotically covariant in the sense of large timelike distances. Without
this property (which usually goes unmentioned) there would be no covariant
scattering theory with an invariant S-matrix and no associated asymptotic Born
probabilities and projection operators. As Born localization goes together
with the quantum mechanical type I$_{\infty}$ algebras, modular localization
is inexorably linked with the monad of QFT. It is fully locally covariant and
permits the formulation of causal propagation over finite distances in terms
of expectation values which however do not permit a further going resolution
in terms of projectors. The absence of probabilities and projectors is related
to the so-called Reeh-Schlieder property namely that the algebra
$\mathcal{A(O)}$ generated by localized fields applied to the vacuum creates a
dense set of states and has no vacuum annihilators.

Ignoring these structural differences produces well-known superluminal
paradoxes as e.g. the alleged causality violation of Fermi's
Gedankenexperiment \cite{Yng}. Although not part of the present investigation,
many of the recent paradoxes in connection with black holes (information
paradox,...) which often are attributed to the still elusive QG probably also
have their origin in an insufficient awareness about the structural
differences between QM and local algebras in QFT.

Some historical remarks are in order. Vacuum polarization as a result of
localization in QFT has been first noticed by Heisenberg when he attempted to
study the well-known Noether connection between global conserved charges and
conserved currents. The Heisenberg infinities in the vacuum fluctuations
caused by the sharp spatial cutoff $R$ can of course be avoided by suitable
testfunction smoothing; in that case the total charge is the $R\rightarrow
\infty$ limit of the partial charge (without changing the testfunction
smoothing at the boundary).

Even more spectacular was the observation of Furry and Oppenheimer that the
application of interacting pointlike fields to the vacuum $\Omega$ does not
only create a one-particle state but also an unavoidable particle-antiparticle
polarization cloud. In modern terminology this observation is part of a
theorem which states that necessary and sufficient for a subwedge
algebra\footnote{This means that the causal closure can be enclosed in a wedge
$\mathcal{O}^{\prime\prime}\subset W.$} $\mathcal{A}\mathrm{(}\mathcal{O})$ to
be generated by free fields is the existence of a PFG (vacuum \textbf{p}%
olarization \textbf{f}ree \textbf{g}enerator, generally an unbounded operator
$G$) affiliated with $\mathcal{A}\mathrm{(}\mathcal{O})$ such that $G\Omega=$ one-particle.

In view of these early perceptions it is a little bit surprising that the
thermal manifestation of modular localization came as late as 1975
\cite{Bi-Wi}. It entered the general consciousness of most particle physicists
only (if at all) through the thermal aspects of black hole horizons
\cite{Haw}\cite{Sew}. Sometimes these thermal manifestations are linked to the
uncertainty relation but this is somewhat misleading. Whereas it is true that
uncertainty relations are behind most phenomena of QM, the thermal
manifestation (KMS states, entropy through localization) of field theoretic
monads $\mathcal{A(O})\subset B(H)$ are outside their quantum mechanical
reign. The entropy related to quantum mechanical entanglement is a "cold" i.e.
information-theoretical kind of entropy whereas the localization entropy is
genuinely thermal.

The rather simple relation between the generators $A_{H(W)}(x_{+},x_{\perp})$
and the original free fields appears much more involved if one asks questions
about inverse holography.

The content of the subsequent sections is as follows. The third section
presents the algebraic setting of lightfront holography in the presence of
interactions. In case of factorizing two-dimensional models for which the
relation between the S-maztrix and the generators of wedge algebras and the
associated chiral holographic projection becomes more explicit. The fourth
section presents two quite different methods to calculate localization
entropy; as expected they agree in their leading behavior in their attenuation
distance for their vacuum polarization at the boundary. Some generic
consequences as the area proportionality of localization entropy and the
necessity to adjust the Bekenstein-Hawking black hole setting to these general
structural facts of QFT before entering the more elusive terrain of QG are
pointed out in the concluding remarks.

\section{Lightfront holography in the presence of interactions}

Since Lagrangian perturbation theory is not really suited for the
investigations of holography and localization thermality, and since there are
presently no reliable nonperturbative construction of models, the passing from
bulk to its holographic projection presently amounts to a purely structural
model-independent discussion. The free field constructions of the previous
section suggests to start the holography from the position of a local operator
algebra monad $\mathcal{A}(W)\subset B(H).$ 

In the interacting situation there is an additional physically stronger
argument\ which is related to the nature of the modular objects of the
standard pair ($\mathcal{A}(W),\Omega$) \cite{AOP}. It is well-known that the
modular group acts as the $W$-preserving boost whereas the anti-unitary
modular inversion $J$ depends on the interaction via the S-matrix and the free
(incoming) modular inversion $J_{0}$%
\begin{equation}
J=J_{0}S_{scat}\label{J}%
\end{equation}
This form of course requires the validity of a complete particle
interpretation which according to S-matrix folklore is expected to follow from
a mass gap. This result (which follows from the TCP invariance of the
S-matrix) shows the close connection between the S-matrix and the monad. which
we want to interpret as a wedge-localized algebra. It can be strengthened by
assuming the validity of crossing "symmetry" for formfactors\footnote{Rigorous
(but unfortunately very elaborate) proofs of crossing only exist for
formfactors with few particles. For two-dimensional factorizing models the
crossing is veryfied as part of the explicite construction.} i.e. the vacuum
to n-particle matrixelements of a localized operator. In that case the inverse
scattering problem can be shown to have a unique solution i.e. if there is at
all a QFT associated to an S-matrix it is necessarily unique (the S-matrix is
the special in-out formfactor of the unitary operator). The physical
significance of the modular objects $\Delta^{it},J$ associated to standard
subwedge localized pairs ($\mathcal{A(O}),\Omega$) is not known apart from the
fact that the action of the modular group is "fuzzy" i.e. cannot be encoded
into a spacetime diffeomorphism. For this reason the wedge algebras play a
prominent role in a new algebraic classification and construction program of
interacting QFTs. Without the use of the simplifications which result from the
holographic projection of the bulk localized in a wedge to its lightfront
horizon such a program would not probably not be feasible. But the fact that a
lot is known about the classification and construction of chiral theories
generates some hope.

The important property of lightfront holography in the context of this paper
is the transverse tensor-factorization for subalgebras \cite{CQG1}\cite{CQG2}
$\left\{  \mathcal{A(O)}\right\}  _{\mathcal{O\subset}LF}$ where $\mathcal{O}$
now stands for regions on $LF$
\begin{align}
&  \mathcal{A(O}_{1}\mathcal{\cup O}_{2}\mathcal{)}=\mathcal{A(O}%
_{1}\mathcal{)\otimes A(O}_{2}\mathcal{)},\text{ }\mathcal{(O}_{1}%
\mathcal{)}_{\perp}\cap\mathcal{(O}_{2}\mathcal{)}_{\perp}=\emptyset
\label{fac}\\
&  \left\langle \Omega\left\vert \mathcal{A(O}_{1}\mathcal{)\otimes A(O}%
_{2}\mathcal{)}\right\vert \Omega\right\rangle =\left\langle \Omega\left\vert
\mathcal{A(O}_{1}\mathcal{)}\left\vert \Omega\right\rangle \left\langle
\Omega\right\vert \mathcal{A(O}_{2}\mathcal{)}\right\vert \Omega\right\rangle
\nonumber
\end{align}
This total absence of transverse entanglement is somewhat surprising. Whereas
this is a rigorous mathematical theorem, the following formulation of
transverse factorization in terms of transverse extended chiral fields
$B_{LF}(x_{+},x_{\perp})$ in analogy to the free field case (\ref{hol}) is
only on the level of a consistency argument%

\begin{equation}
\left[  B_{LF}^{(i)}(x_{+},x_{\perp}),B_{LF}^{(k)}(x_{+}^{\prime},x_{\perp
}^{\prime})\right]  \simeq\left\{  \sum_{l}\delta^{n_{l}}(x_{+}-x_{+}^{\prime
})B^{(l)}(x_{+},x_{\perp})\right\}  \delta(x_{\perp}-x_{\perp}^{\prime})
\label{c.r.}%
\end{equation}
where in the case of transversely extended rational theories the algebraic
structure of the theory permits a characterization in terms of a finite number
of $LF$ generating fields $B_{LF}^{(i)}.$ The transverse delta functions
result from transverse derivatives in the $B$-fields; they are associated to
non-fluctuating quantum mechanical degrees of freedom since in contrast to the
lightlike positive energy condition they suffer no frequency restriction and
hence appear already on the level of correlation functions. This commutation
relation certainly holds for Wick-monomials of free fields. Note however that
unlike the free field case (\ref{hol}), in the presence of interactions one
should not expect a factorization into longitudinal and transverse part i.e.
the $B^{\prime}s$ will remain $x_{\perp}$ dependent, i.e. holography leads to
a (transversely) \textit{extended chiral theory}.

In the presence of interactions one expects the appearance of fields with
anomalous dimensional (non-integer dimensional but bosonic) fields in the
bulk. Whereas in the bulk the anomalous dimension is independent of its
bosonic spacelike commutativity, in (extended) chiral theories the
spin-statistics relation together with the relation between spin and scale
dimension excludes bosonic observables with anomalous scale dimension. This
indicates that the global algebra which the compactly localized subalgebras
algebras of $\mathcal{A}(LF)$ and $\mathcal{A}(\partial W)$ generate maybe
smaller than these algebras i.e.%
\begin{equation}
\overline{%
%TCIMACRO{\dbigcup \limits_{\mathcal{O}\subset\partial W}}%
%BeginExpansion
{\displaystyle\bigcup\limits_{\mathcal{O}\subset\partial W}}
%EndExpansion
\mathcal{A(O})}\subseteq\mathcal{A}(\partial W)\equiv\mathcal{A}(W)
\label{alg}%
\end{equation}
We will return to this important problem in the more restricted context of
d=1+1 factorizing models.

Whereas in higher spacetime dimensions the pointlike generating property of
algebraic nets still depends on certain technical assumption, it is well-known
that the simpler chiral algebraic nets always have pointlike generators
\cite{Joerss}. Looking at the representation theoretical nature of the
argument there can be no doubt that conformal nets in higher spacetime
dimension also share this pointlike generator property. For massive
factorizing models these fields are known in terms of infinite series whose
convergence status is not yet known.

Many profound ideas of the old bootstrap of the 60s were lost as a result of
its ideological hubris which led to a positioning of the S-matrix bootstrap
against QFT instead of considering the use of on-shell objects and their
properties (e.g. the crossing property) as a valuable enrichment of QFT. In
the new context of a mass-shell based construction of QFT which tries to
extend Wigner's representation theoretical approach for one-particle spaces to
the realm of interactions some of these old bootstrap ideas come to new life.
A pure bootstrap approach in which the classification of interacting
S-matrices can be pursued separated from the construction of an associated QFT
is limited to two spacetime dimensions; in higher dimensions the S-matrix
plays the role of a special formfactor and the formfactor program in turn
becomes incorporated into the construction of generators for wedge algebras.

\section{The special case of factorizing models and their holographic
projection onto a chiral QFT}

Many profound ideas of the 60s, which were lost in the aftermath of the "TOE
hubris" when the S-matrix bootstrap program was positioned against instead of
within QFT, were later on vindicated when it was observed\footnote{It was
based on the integrability features oberserved in quasiclassical approximation
of the particle spectrum in certain two-dimensional QFTs by Dashen, Hasslacher
and Neveu which in case of the Sine-Gordon theory was afterwards explained in
terms of the bootstrap S-matrix properties restricted to two-particle elastic
scattering \cite{STW}.} that there is a rich class of interacting
two-dimensional massive QFT which are uniquely associated with bootstrap
S-matrices. The first observation which led to a kind of revival within a more
limited context of integrable QFT was about the possible integrability of
certain two-dimensional Lagrangian QFT on the basis of their quasiclassical
mass spectrum. This was followed by the remark that in case of the Sine-Gordon
model the mass formula follows from an exact computation based on applying the
bootstrap requirements to an elastic two-particle S-matrix Ansatz \cite{STW}.
In two dimensions it is consistent with macro-causality (in particular with
the cluster factorization property) to have a purely elastic S-operator which
factorize in terms of an elastic two-particle scattering amplitude; the
associated QFTs were therefore were named \textit{factorizing models}. These
S-matrices were susceptible to a systematic classification obtained from the
bootstrap principles: unitarity, Poincar\'{e} invariance and the crossing
property \cite{Ka1}. This bootstrap S-matrix setting was then connected via
the bootstrap-formfactor program with a new way of nonperturbative
construction of factorizing QFTs \cite{KW}.

Different from the original hope that the bootstrap principles would lead to
unique TOE\footnote{Apart from gravity which was already missing in the
Heisenberg's "Weltformel", probably the first futile attempt at a TOE before
this strange ideas like that became fashionable in particle physics.}, it
turned out that the factorizing setting rather led to a extraordinary rich
family of infinitely many nontrivial two-dimensional nonperturbatively
controllable models. Besides those models which permit a Lagrangian
description (in the sense of a Lagrangian "baptism" and a divergent
perturbative series rather than a construction), there is an enormous number
of non-Lagrangian models (the oldest and most prominent being the $Z(N)$ model
.\cite{Ba-Ka}) which is a consequence of the fact that there are more elastic
two-particle bootstrap S-matrices $S^{(2)}$ (i.e. more solution of the
Yang-Baxter equations) than interactions $\mathcal{L}_{int}.$ There is no
physical principle which distinguishes Lagrangian QFT relative to those of
non-Lagrangian origin since the quantization parallelism to classical physics
is not a physical principle and there is no intrinsic autonomous property of
QFT which is capable to reveal a Lagrangian origin.

The next step was to encode the elastic S-matrix data into an algebraic
structure. The resulting Zamolodchikov-Faddeev (Z-F) algebra has the
appearance of a non-local generalization of the Wigner momentum space
creation/annihilation operators into which the S-matrix data enter as
structure coefficients which characterize the commutation structure
(\ref{Zam}). The observation that the on-shell Fourier transforms into x-space
yields a covariant on-shell generator of a wedge-localized algebra bestows a
\textit{physical interpretation in terms of spacetime localization} to these
initially purely auxiliary nonlocal operators \cite{AOP}. Modular localization
and the related Tomita-Takesaki modular operator theory of operator algebras
plays a crucial role in obtaining these results.

As mentioned in the previous section modular theory also shows that, contrary
to the old bootstrap philosophy, the S-matrix of a QFT is not completely void
of spacetime localization aspects. Rather it is deeply connected with
wedge-localization in the sense that S is a \textit{relative modular
invariant} of the wedge-algebra (\ref{J}). Considered as a in-out formfactor
of the unit operator, and combined with the formfactors of all other
operators, the S-matrix leads to the \textit{uniqueness} of an associated QFT
(if it exists) \textit{if the formfactors fulfill the crossing property}
\cite{uniqueness}. In general crossing only assures the uniqueness of the
inverse scattering problem; no argument is presently known which secures its
existence in a general setting including crossing.

Crossing is a deep analytic on-shell property which is not valid in other non
QFT based relativistic S-matrix theories which implement interactions directly
(without the mediation of QFT) in a Wigner multiparticle representation
theoretical setting, the so-called \textit{direct particle interactions}
\cite{C-P}\cite{interface}. Hence crossing may be considered as the on-shell
imprint of the field theoretic locality and spectral properties in conjunction
with the assumption of completeness of asymptotic particle configurations. In
the nonperturbative setting of QFT it was only proven for special particle
configurations; but since in factorizing theories crossing is part of the
construction, the existence proof for those models also shows the validity of
this property within this class of factorizing QFTs.

The setting of two-dimensional factorizing models is presently the only known
case in which the bootstrap-formfactor conjecture (S-matrix, formfactor)
$\rightarrow$ QFT can be backed up by constructive mathematical steps. This
class of factorizing models is not only interacting in the global sense of a
nontrivial S-matrix, but it also permits a local characterization in the sense
of possessing no subwedge localized $\mathcal{A(O})$-affiliated PFGs
(vacuum-\textbf{p}olarization-\textbf{f}ree-\textbf{g}enerators\footnote{PFGs
are operators which applied to the vacuum generate a one-particle state with
admixture of vacuum polarization clouds. Only for free fields they exist for
compact localization regions.}). The different type of interactions in the
latter case would instead of corresponding to different Lagrangians be
characterized by different S-matrices (different wedge generators) or on the
local level by different shapes of the local vacuum polarization clouds (still
somewhat futuristic). These properties are expected to continue to hold also
in the general case when there are no so called \textit{temperated
wedge-localized PFGs} (see below).

The characteristic feature of factorizing models, which entails the relative
simplicity of their construction, is that they possess \textit{no real
(on-shell) particle production}. But the characteristic feature of interacting
QFT is not particle production (which can also be incorporated into
\textit{direct particle theories \cite{C-P}}) but rather the
\textit{interaction-caused infinite vacuum polarization clouds} which result
from compact spacetime localization. This is in contrast to the localized
interaction-free algebras which always admit PFGs leading to the existence of
underlying free field generators for arbitrary localizations and for which the
composite fields (Wick-powers) generate vacuum polarization with only a finite
number of particle/anti-particle pairs (the number depending on the degree of
the Wick monomial). The existence of temperate PFGs for wedge regions only
maintains the on-shell wedge generators cIose to their free field form; they
obey the slightly more general Z-F commutation relations which lead to the
factorization of the S-matrix which is the origin of the terminology for this
class of models.

A systematic structural study of PFGs in the presence of interactions in
general QFT was initiated in \cite{Bo-Bu-Sc} and it was found that whereas for
compact localization regions any operator has the infinite vacuum polarization
cloud (confirming previous results), it is only for the noncompact wedge
region where modular theory leads to the existence of PFG.

Unfortunately modular theory does not guaranty reasonable domain properties
which permit a successive application of these unbounded operators as for
smeared Wightman fields. Only if these wedge-localized PFGs are
\textit{temperate} \cite{Bo-Bu-Sc} in the sense of good domain properties one
has been able to extract interesting consequences from their existence.
According to an old theorem \cite{Aks} this is impossible in higher
dimensions; using some rigorous analytic properties for the scattering
amplitude one finds that this is only possible in d=1+1 \cite{Bo-Bu-Sc}. With
the help of crossing for formfactors one can exclude higher direct elastic
multiparticle amplitudes so that the higher multiparticle scattering has to go
through $S^{(2)}$ which is the S-matrix definition of factorizing QFTs.

So the notion of integrable or factorizable theories can be fully substituted
by the notion of \textit{theories with temperate PFG's}. With this change of
paradigmatic emphasis, QFT returns to its beginnings when Furry and
Oppenheimer discovered (perturbatively) the omnipresence of interaction-caused
vacuum polarization clouds; but now with rich additional conceptual additions
from modular localization theory and the hope that perturbative Lagrangian
quantization can be replaced by characterizing interacting theories in terms
of wedge generators and structural properties of subwedge polarization as it
is presently happening in factorizing models.

Modular operator theory applied to the wedge-localized algebra of QFT
$\mathcal{A}(W)$ leads to a new semilocal interpretation of S-matrices: as
mentioned before (\ref{J}) they are \textit{relative modular invariants}
(relative to the free wedge algebra $\mathcal{A}_{0}(W)$ without
interactions). This important insight which links scattering to fields (i.e.
part of the inverse scattering program) was missing in the old S-matrix
bootstrap approach. It confirms that the S-matrix preempts an important
semiinfinite localization aspect of QFT and that it was not wise to position
the S-matrix bootstrap program \textit{against} QFT not to mention the aborted
attempt to convert it into a TOE.

The for the time-being last step in this interesting sequence of events is
Lechner's recent existence- and particle completeness- proof \cite{Lech} in
the setting of factorizing models. It is based on the phase-space property of
\textit{modular nuclearity;} this result in some way vindicates the four
decades old Haag-Swieca paper on the connection of local fields and asymptotic
particles via phase space properties by giving a positive answer to the
question in the title of their paper. It also shows that particle physics does
not only consists of changing fashions but that there are once in a while
ideas with a historical breath.

Let us sketch some details of these constructions. The starting point is the
formula for the wedge generators in terms of the Z-F algebra operators. In the
scalar one-component case these generators have a form similar to free field,
except that the on-shell creation/annihilation operators which appear in their
Fourier transform fulfill the Z-F commutation relation\footnote{The $\tilde
{Z}^{\#}(\theta)$ in general are multi-component and the *-algebra requirement
forces the matrix-valued structure functions $S(\theta)$ algebra to be unitary
solutions of the Yang-Baxer equations. It follows in the course of
construction of the theory that the S-coefficients of the algebra are also the
two-particle matrixelements of the scattering matrix (showing again the close
relation of the scattering matrix to local aspects of QFT).} (which are
slightly more general than the Wigner creation/annihilation operators).%

\begin{align}
Z(x)  &  =\frac{1}{\left(  2\pi\right)  ^{\frac{1}{2}}}\int(e^{ip(\theta
)x}\tilde{Z}(\theta)+h.c)\frac{d\theta}{2},\text{~}p(\theta)=m(ch\theta
,sh\theta)\label{Z}\\
&  \tilde{Z}(\theta)\tilde{Z}^{\ast}(\theta^{\prime})=S(\theta-\theta^{\prime
})\tilde{Z}^{\ast}(\theta^{\prime})\tilde{Z}(\theta)+\delta(\theta
-\theta^{\prime})\label{Zam}\\
&  \tilde{Z}(\theta)\tilde{Z}(\theta^{\prime})=S(\theta-\theta^{\prime}%
)\tilde{Z}(\theta^{\prime})\tilde{Z}(\theta)\nonumber
\end{align}
This has the consequence that although the $\tilde{Z}^{\#}$ commutation
relations remain close to those of the Wigner-Fock creation/annihilation
operators and the field $Z(x)$ transforms like a would be pointlike field
under the Poincar\'{e} group, the $Z(x)$ is \textit{not pointlike local }\ in
the sense of spacelike commutation relations. It also shows a covariant
coordinate label $x$ is not necessarily indicating pointlike localization,
although in the opposite direction the statement is correct.

On the other hand $Z(x)$ is \textit{not completely nonlocal} either; since the
application of modular operator theory reveals that operators $Z(f)$ with
$suppf\in W$ are only $W-$local (\textit{semilocal),} no matter how sharply
one localizes the support inside $W.$ The exponentiation of unbounded smeared
$Z(f)$ operators leads to the well known Weyl algebra structure whose weak
closure (or double commutant) defines the operator algebra $\mathcal{A}(W)$
which according to modular theory is W-localized.

In this way the Z-F operators, which at first appeared as purely formal
auxiliary objects in the formfactor-bootstrap program, are given a spacetime
interpretation. They are objects which have better relative locality
properties (always relative to the pointlike interacting fields) than the
in-out free fields. Instead of being completely relatively nonlocal they at
least are relative wedge local with respect to the interacting fields and with
respect to themselves and hence fully wedge local. The nonlocality (better
semilocality) of $Z(x)$ is the prize to pay for the absence of the vacuum
polarization in the vector-valued distribution $Z(x)\Omega.$

It is not our aim here to study nonlocal theories for their own sake; it is
the fact that they are \textit{semilocal in the sense of wedges} which makes
the $Z(x)$ very interesting as intermediate objects on the way to genuine
local theories.

This has an interesting connection with a kind of the particle--field
complementarity in the presence of interactions\footnote{It should not be
confused with Bohr's particle-wave complementarity which persists without
interaction.}. In a free theory one-particle states can be generated from the
vacuum by the application of smeared fields with smearing functions of
arbitrary small support (PFG's exist for arbitrary small localization
regions). The presence of any interaction radically changes this state of
affairs \cite{interface}: there is no relative compact region for which the
PFG property prevails and the smallest noncompact causally complete region
(causal completion is automatic in the algebraic approach) for which PFGs
prevail even in the presence of interactions are the wedges. In this sense the
\textit{wedge region leads to the best compromise between particles and
fields}.

Particles are not only important in scattering theory\footnote{Scattering
theory is build on the idea that multiparticle states asymptotically
stabilize: if at asymptotically large times by counter coincidence and
anticoincidence arrangements one established the presence of an precisely
n-fold localized state then it remains this way.} but they play the crucial
role in nonperturbative construction of models via the bootstrap-formfactor
construction of factorizing models and the relative modular invariance
property for wedges of the S-matrix. In contradistinction to quantum fields
which coordinatize algebras (similar to the role of coordinates in the modern
formulation of differential geometry) and are therefore not of \textit{direct}
physical significance, \textit{particles have an ontological
individuality/objectivity} \cite{interface}. But in the presence of
interactions they are in the above sense nonlocal, and therefore it is not
surprising that nonlocal intermediate steps are helpful in nonperturbative constructions.

Schematically one proceeds as
\begin{equation}
temperate~PFG\text{ }for\text{ }wedge\longrightarrow ZF-algebra\longrightarrow
\mathcal{A}(W)
\end{equation}

One-particle states free of vacuum-polarization can always be created from the
vacuum by the application of unbounded wedge-localized PFGs \cite{Bo-Bu-Sc},
which permits a particle interpretation. But in but in order to be able to
utilize them for constructive purposes one must presently require that they be
"temperate", i.e. their range is tuned to their domain in such a way that an
iterative applications (as in Wightman field theory) is possible. It is
precisely this requirement which forces the restriction of factorizability and
therefore the two-dimensionality QFT on us and it is only in this special
setting that the bootstrap classification and computation of S-matrices can be
separated from the formfactor construction which requires the setting of QFT.

In view of the fact that historically the first investigations of factorizing
models proceeded through the quantization of classically integrable field
models, and in view of the complicated nature of classical integrability
(infinitely many conservation laws), it comes as a pleasant surprise that a
simple restriction in terms of vacuum polarization leads to the same result in
a purely intrinsic QFT way. It is one of several known instances in which
quantum arguments are conceptually simpler than their classical counterparts.

The construction of the wedge algebra $\mathcal{A}(W)$ from the $Z$-generators
is entirely analogous to the construction of the Weyl algebra from the free
fields. Whereas on the level of the S-matrix and the wedge generators of the
form \ref{Z} the theory has the appearance of a relativistic potential theory,
this state of affairs changes radically if one passes to compact localizations
as the Poincar\'{e} covariant family of double cone algebras $\mathcal{A(C})$
which arise as relative commutants of wedge algebras%
\begin{equation}
\mathcal{A(C})=%
%TCIMACRO{\dbigcap \limits_{W\supset\mathcal{O}}}%
%BeginExpansion
{\displaystyle\bigcap\limits_{W\supset\mathcal{O}}}
%EndExpansion
\mathcal{A}(W)
\end{equation}
where in the two-dimensional case one only needs to intersect the standard
wedge (with apex at the origin) with its translated opposite. Intersecting
algebras is a task for which presently no tools exist. However, as a result of
the simplicity of the algebraic structure of factorizing models, there two
ways to do this. The more formal procedure starts from a general Ansatz
($p(\theta)=m(ch\theta,sh\theta)$)%
\begin{equation}
A(x)=%
%TCIMACRO{\dsum }%
%BeginExpansion
{\displaystyle\sum}
%EndExpansion
\frac{1}{n!}%
%TCIMACRO{\dint _{C}}%
%BeginExpansion
{\displaystyle\int_{C}}
%EndExpansion
d\theta_{1}...%
%TCIMACRO{\dint _{C}}%
%BeginExpansion
{\displaystyle\int_{C}}
%EndExpansion
d\theta_{n}e^{-ix\sum p(\theta_{i})}a(\theta_{1},...\theta_{n})\tilde
{Z}(\theta_{1})...\tilde{Z}(\theta_{1}) \label{onshell}%
\end{equation}
where for reasons of a compact notation we view the creation part $\tilde
{Z}^{\ast}(\theta)$ as the $\tilde{Z}(\theta+i\pi)$ i.e. as the Z on the upper
boundary of a strip\footnote{The notation is suggested by the the strip
analyticity coming from wedge localization. Of course only functions but not
field operators or their Fourier transforms can be analytic.} (we could have
introduced this notation already in (\ref{Z})).

This is similar to the GLZ representation of the interacting Heisenberg field
in terms of incoming free field, in which case the spacetime dependent
coefficient functions turn out to be on-shell restrictions of Fourier
transforms of retarded functions except that instead of the on-shell incoming
fields one takes the on-shell $Z$ operators which conceptually are somewhere
between Heisenberg and incoming fields.

In the latter case the coefficient functions are precisely those formfactors
which feature in the bootstrap-formfactor approach to factorizing theories.
\ Together with a certain analytic requirement on the coefficient functions,
the space of these formal power series represent the space of formal (in the
spirit of vertex operators) W-localized fields. Taking for $\mathcal{O}$ a
double cone $\mathcal{D}$ whose left apex coalesces with the origin and
representing $\mathcal{D}$ as the intersection of the standard right wedge
with an $a>0$ translated standard left wedge, the calculation of
$\mathcal{A(D})$-generating operators is based on the relative commutant
restriction placed on the coefficient functions:%
\begin{equation}
\left[  A(x),U(a)U(j)Z(f)U(j)^{\ast}U(a)^{\ast}\right]  =0
\end{equation}
In words: the generators of the right wedge (the $A^{\prime}s$ whose
coefficient functions have the correct $\theta$-strip analyticity
corresponding to the right wedge localization) are subjected to the
restriction that only those which commute with the generating Z's of the
$a$-shifted opposite wedge are admitted (i.e. the ones which generate
$\mathcal{A}(\mathcal{D})).$ Here $U(j)$ is the free TCP operator i.e. the one
which acts on the multiparticle wave functions in the standard way and
therefore $Z^{opp}(f)=U(j)Z(f)U(j)^{\ast}$ are generators of the opposite
wedge. Since the commutation with the restricting operators $Z_{a}^{opp}$ map
the $n^{th}$ order term in $A$ with the adjacent n+1 and n-1, one obtains a
rather simple linear recursion for the coefficient functions.

In praxis one uses this commutation relation together with covariance in order
to construct a basis of composite fields within each superselection sector.
Formally the space of generating operators\footnote{For unbounded operators
associated to algebras (of bounded operators) it is more appropriate to speak
about spaces than algebras if one has not said anything about dense domains.}
for compactly localized operators is given in terms of infinite series in the
Z operators with coefficient functions which obey the same relations which are
known as "the formfactor axioms" in the bootstrap formfactor approach
\cite{Smirnov}. One obtains an infinite space of field generators in terms of
the infinite space of formfactors in the bootstrap-formfactor program.

As in the case of free fields and their Wick composites one has "basic" (there
is no Lagrangian hierarchy here) fields which by definition are those
pointlike fields which if together successively applied to the vacuum generate
the Hilbert space (they act cyclically on the vacuum) and the remainder are
composites. The latter are expected to look like classical local monomials in
the basic fields except that there is a spacetime limiting \textit{normal
order prescription}. In the Z-expansion it is easy to see that the composites
share with the basic fields the nucleus of the formfactor construction
(minimal formfactor) and deviate only in certain momentum space polynomials,
but to translate these observations into normal product formulas for
composites is not possible within the present state of QFT technology.

Since attempts to show convergence of (\ref{onshell}) have failed\footnote{It
is perfectly consistent (with everything which one knows about divergence of
perturbative series) that these series diverge; since each single term is
analytic in a small circle around zero coupling, this would put the blame of
divergence of perturbative series of off-shell objects on vacuum
polarization.}, it is deeply satisfying that there is at least an existence
proof for nontrivial intersections of wedge localized algebras based on
phase-space behavior ("nuclear modularity") \cite{Lech} which allows to bypass
the convergence problem of such series representations. This shows that the
algebraic setting is not only a valuable conceptual
field-coordinatization-free guide to the get to the right starting point for
doing calculations in terms of field coordinates (which is the way we used
it), but that it is also capable to shed some new light on age old problems of
QFT, as the problem of their existence beyond formal perturbative power series
with the unresolved convergence status.

The on-shell representation of Heisenberg fields (\ref{Z}) as an infinite
series is a particularly useful starting point of the holographic projection
since apart from the convergence problem it is mathematically and conceptually
less demanding. It avoids the use of the still somewhat unfamiliar modular
theory and uses the more standard apparatus of QFT. Instead of aiming at
rigorous proofs it satisfies itself with consistency checks. and delegates the
more ambitious existence proofs based on modular theory to a second stage of
mathematical refinement.

With the help of the infinite series expansion (\ref{onshell}) we can proceed
along the lines of a naive restriction argument (restricting plane wave
factors to a lightfront) as was done for the free field by using its on-shell
representation. But there is one stain which should not be supressed In order
to arrive at on-shell formulas one has to go through the nonlocal steps of
scattering theory. Hence the use of such on-shell formulas is somewhat against
the spirit of simplification of certain properties (as the short-distance
behavior) through lightfront holography \textit{before} starting any calculation.\ 

Let us take two well studied models and extract some interesting informations
from their two-point function using their holographic series representation.
According to the previous remarks, the general formfactor series
representation for the holographic two-point function reads (x$_{+}$
translational lightray variable, $p_{-}(\theta)=e^{-\theta}$)%
\begin{align}
w(x_{+})  &  =1+%
%TCIMACRO{\dsum }%
%BeginExpansion
{\displaystyle\sum}
%EndExpansion
\frac{1}{n!}%
%TCIMACRO{\dint }%
%BeginExpansion
{\displaystyle\int}
%EndExpansion
d\theta_{1}...%
%TCIMACRO{\dint }%
%BeginExpansion
{\displaystyle\int}
%EndExpansion
d\theta_{n}e^{-ix_{+}\sum p_{-}(\theta_{i})}b(\theta_{1},...\theta_{n})\\
&  =\exp%
%TCIMACRO{\dsum }%
%BeginExpansion
{\displaystyle\sum}
%EndExpansion
\frac{1}{n!}%
%TCIMACRO{\dint }%
%BeginExpansion
{\displaystyle\int}
%EndExpansion
d\theta_{1}...%
%TCIMACRO{\dint }%
%BeginExpansion
{\displaystyle\int}
%EndExpansion
d\theta_{n}e^{-ix_{+}\sum p_{-}(\theta_{i})}b_{c}(\theta_{1},...\theta
_{n})\label{dim}\\
&  b(\theta_{1},...\theta_{n})=\left\vert a(\theta_{1},...\theta
_{n})\right\vert ^{2}\nonumber
\end{align}
where in the second line we have used the Ursell-Mayer expansion which
expresses the coefficient functions in terms of their cumulants $b_{c}$.
Obviously the $lnw$ series is more convenient if we are interested in the
anomalous dimension of the holographically projected field.

From this series one may read off the anomalous dimension of the respective
field. Apart from the critical limit in the work of Babujian and Karowski
\cite{Ba-Ka} which in the holographic approach is replaced by an exact
bulk-boundary relation (and not by another bulk theory in the same
universality class), we can take over all their formulas, in particular the
formula for the dimension $d_{A}$ of a algebra-affiliated field $A(x)$%
\begin{align}
w(x_{+})  &  =const\left(  x_{+}\right)  ^{-2d_{A}}\label{anom}\\
d_{A}  &  =\frac{1}{2}%
%TCIMACRO{\dsum }%
%BeginExpansion
{\displaystyle\sum}
%EndExpansion
\frac{1}{n!}%
%TCIMACRO{\dint }%
%BeginExpansion
{\displaystyle\int}
%EndExpansion
d\theta_{1}...%
%TCIMACRO{\dint }%
%BeginExpansion
{\displaystyle\int}
%EndExpansion
d\theta_{n-1}b_{c}(\theta_{1},...\theta_{n-1},0)\nonumber
\end{align}
For the Ising field theory one can do all the integrals as in \cite{Ba-Ka} and
then sum the series in order to obtain the expected result $d_{A}=\frac{1}%
{16}.$ For the Sinh-Gordon model the contribution to the series up n=2, the
authors arrive at a rather complicated function in terms of the Sinh-Gordon
coupling strength whose further evaluation has to be done numerically.
Holography with pointlike fields leads to the same integrals, its only
advantage is that its relation to the bulk is exact since it does not change
the algebraic substrate but only its spacetime ordering.

The main difference to the present derivation is conceptual; whereas Babujian
and Karowski go to the critical limit which is associated with a massless QFT
associated with different operators which act in a separate Hilbert space,
holography takes place in the same Hilbert space and certain wedge-localized
algebras whose upper causal horizon lies on the lightfront are shared between
bulk and holographic projection. The set of shared algebras is invariant under
a certain 2-parametric subgroup of the 3-parametric Poincar\'{e} group
$\mathcal{P}$ and in order to re-construct the mass spectrum one must know how
the missing Poincar\'{e} transformation (e.g. the opposite lightray
translation) acts on these shared subalgebras. The action on $LF$ is
necessarily nonlocal (fuzzy) i.e. it cannot be described in terms of geometry.

On the other hand the holographic projection acquires a new symmetry whose
presence was not noticed in the bulk description, namely the Moebius rotation
which together with the two transformations inherited from the bulk
constitutes the 3-parametric Moebius group $SL(2,R).$ The reason why it was
not noticed in the bulk is because its action on the bulk is "fuzzy"; only on
the horizon it becomes geometric. The concept of algebraic transformations of
the bulk which become which are not related to Noether 's theorem and become
only geometric upon restriction is a new not yet explored structure of QFT.

In fact one expects the holographic projection to have the covariance under
the full diffeomorphism group $Diff(S^{1}),$ even though it does not arise
from a chiral decomposition of two-dimensional conformal QFT$.$ The beauty of
factorizing models is that as a result of the presence of the Z-F algebra one
can study all these questions in a reasonably controllable setting i.e.
\textit{factorizing models are presently the best theoretical laboratory for
testing conjectures} beyond perturbation theory.

Among the ideas waiting for a test is the conjecture that not only the free
field holography for which the Diff-invariance is an obvious consequence of
the transverse-longitudinal factorization of the two-point function
(\ref{hol}), but also the holographic projections of factorizing models are
automatically Moebius invariant and (under mild additional restrictions) even
Diff(S$^{1}$)-covariant. In higher dimensions the invariance group is expected
to be even larger in the sense that the algebraic holographic structure also
allows certain $x_{\perp}$\textit{-dependent chiral diffeomorphisms }which
are\textit{ }automorphisms of the algebraic commutation structure of extended
chiral theories.

The fact that the holographic projection has more symmetries than those of the
invariance group of the lightfront has been called symmetry enhancement on the
horizon \cite{Su-Ve}. In the case of the Moebius rotation within factorizing
models it means in particular that the rotation generator $L_{0}$ can be
written as an infinite series in the $Zs,$ whereas the translation and
dilation retain their usual bilinear form. Besides the commutation relations
the $L_{0}$ is restricted by the requirement that the vacuum is annihilated.
Since the one particle creation in the bulk looses its physical meaning in the
holographic projection, the application of $L_{0}$ to the one-Z state
$Z(x)\Omega$ adds infinitely many Z-"quanta". In other words the Z-description
is not a very natural basis if used within in a chiral theory since it has
these unusual aspects.

Another important structural problem which still awaits clarification is the
question how the more rigorous algebraic holography (for details we refer to
\cite{CQG1}\cite{CQG2}\cite{interface}) is related to the holographic
projection in terms of pointlike fields. The obvious conjecture in case of
factorizing models is that the holographic bosonic observable algebras are
generated by the holographic projection from bulk field which in addition of
being bosonic also have integer short distance dimension. Here the free field
is atypical because in that case all composites have integer dimension and
there are no bulk fields with anomalous which survive the algebraic holography
process which only passes the those operators which are bosonic in the sense
of the lightray and which therefore must have integer scale dimension on the lightray.

On the other hand the holographic projection in the sense of pointlike fields
does not suffer these restrictions to integer short distance behavior in the
bulk, but those anomalous dimensional bulk fields will loose their bosonic
spacelike commutation structure upon holographic restriction and have
braid-group commutation relations on the lightray. So in case of algebraic
holography for factorizing theories there seems to be no alternative than to
reconstruct the missing plektonic fields via the DHR superselection theory.

Most of the statements and conjectures, except those involving $Zs$ can be
formulated in higher dimension. The higher the spacetime dimension, the more
lightfront changing transformation one must apply in order to recover the
local structure of the bulk from that of the lightfront by inverse holography.

The holographic projection is an excellent method for calculating properties
which are caused by the spacetime localization of quantum matter \ as e.g. the
entropy of localization. Since this entropy results from the infinite vacuum
polarization cloud on the boundary \ of localization, it is not necessary to
know details about the localization substructure inside the bulk. This
legitimizes to perform entropy calculations in the holographic projection i.e.
to reduce the calculation of localization entropy for the wedge algebra to
that for a semi-line which is conformally equivalent to an interval.

We know since Heisenberg's times that the vacuum polarization of sharply
localized relativistic matter is infinite and therefore we have to
attenuate\footnote{In the case of "partial" charges which are formally
obtained by integrating the zero compontent of a quantum current over the
volume V this is done by smearing with a test function which goes to zero
smoothly within a finite "attenuation collar" which is attaches to the volume.
In the infinite volume limit the dependence on the smearing function drops out
and one obtains the global charge.} those particle/anti-particle pairs by a
"split procedure" (see next section) which requires to approximate the
interval from the inside by a sequence of smaller intervals. Conceptually this
is not much different from the formation of the thermodynamic limit for a
heat-bath thermal theory. The prerequisite for this relation is that the
global algebra in the heat bath representation defined by a KMS state and the
global algebra in the vacuum representation after restriction to a localized
subalgebra are of identical type. This is the case since both algebras are of
the same type, they are what we called a \textit{monad} in \cite{interface} As
a result of the conformal invariance after holographic projection, even the
geometric description becomes conformally equivalent; the infinite volume
factor (i.e. infinite length $l$) of the in the holographic lightray theory is
to be replaced by the logarithm of a diverging invariant $\varepsilon$ which
one can form from 4 points and which goes to zero as the shortest distance of
the endpoint of the smaller to those of the bigger interval the distance i.e.
diverges as $\left\vert ln\varepsilon\right\vert $ The result which will be
derived in the next section.

There is one more reason why the holographic projection is the preferred
method for dealing with bulk properties in particular in the case of
factorizing models. The Z-generators (\ref{Z}) of the half-lightray are the
same as those for the wedge, except that the plane wave factors are those of a
one-dimensional QFT. Instead of determining operator algebras associated with
intervals on lightlike lines via algebraic intersections and derive the
Moebius covariance via modular theory, one can also try to find a formula for
the \textit{Moebius rotation in terms of a series in the Z-operators.}

One expects that the Moebius invariance continues to be valid beyond the
holographic projection of the free theory. The convergence of the infinite
series which represents the anomalous dimension (\ref{dim}) of the holographic
projection of the bulk disorder field in the massive Ising model to the
correct value for the chiral Ising model is an encouraging consistency check.

The formal arguments in favour of the Moebius covariance of holography are on
the same level of rigor as the "proof" of dilation invariance of the zero mass
limit, but here we want to stay close to the spirit of mathematical physics
where consistency checks are not sufficient.

It is very important to understand these connections between bulk and its
holography, and the factorizing models provide presently the best theoretical
testing ground. For people who know chiral QFT via the standard approach, it
is highly surprising that such theories (at least in those cases where they
arise via holographic projections) have another (in addition to the $L_{0}$
Fourier decomposition) particle-like description in terms of a non-Moebius
covariant Z-system (it lacks rotational Moebius covariance since the presence
of Z-polarization clouds causes a complicated transformation property under
Moebius transformations).

Whereas several results in this section depend on the factorizability of the
model, the idea that the structure of the wedge algebra should form the
central spine of a new completely intrinsic constructive approach to QFT is
generic. Naturally nobody with any experience in particle physics would
expects that outside of factorizing models one can calculate an S-matrix
exactly using only the bootstrap prescription. Since the S-matrix in the
present setting is the formfactor of the identity operator, on should rather
view the determination of the S-matrix as part of the formfactor program where
all formfactors must be determined together. The crucial hint comes from
modular theory which relates the S-matrix to the so-called modular inversion
(\ref{J}) which coalesces (apart from a spatial rotation) with the TCP operator.

This permits to think about an onshell perturbative approach for formfactors
in which the interaction input is not a Lagrangian but rather a lowest order
S-matrix . Since in such an approach there is no place for singular pointlike
fields but only for generators of wedge algebras, one does not expect new
parameters arising from renormalization Hence in such a still futuristic
perturbative setting there should be many more finite parametric models
invariant under renormalization group transformation than in the pointlike
Lagrangian renormalization approach. Such an explosion of new finite
parametric models is already evident in the factorizing situation where e.g.
the infinitely many Sinh-Gordon type S-matrices which one obtains via CDD pole
modifications all have uniquely related QFT but no Lagrangian name \cite{Lech}.

A very interesting soluble theory (as a result of its exotic statistics and
crossing relation) is the Z(N) model. It derives its name not from a
Lagrangian (In fact for N%
%TCIMACRO{\TEXTsymbol{>}}%
%BeginExpansion
$>$%
%EndExpansion
2 it is not expected to have any). but rather because it was defined by the
requirement that its S-matrix should implement the idea: antiparticle = bound
state of N-1 particles Which is the minimal way of implementaing nuclear
democracy within Z(N) symmetry. There is no reason to believe that QFT is
tight to "baptizations" in higher dimensions.

\section{The area density of localization-entropy via the inverse Unruh
effect}

After having established the d-2 dim. area proportionality of localization
entropy, the remaining task is to use the rather detailed knowledge about
chiral theories in order to calculate the dependence of this area density on
the variable attenuation size $\varepsilon$ of the vacuum polarization cloud.

There are two quite different ways to achieve this. One is based on a kind of
\textit{inverse Unruh effect} for chiral theories: the monad $\mathcal{A}%
(0,\infty)$ with respect to the vacuum is unitarily equivalent (via a
conformal map) to a KMS state at $T=2\pi~$on the global algebra $\mathcal{A}%
(-\infty,+\infty),$ in terms of the standard pair notation (the halfcircle
after the compactification $\mathbb{\dot{R}}=S^{1}$)%
\begin{equation}
(\mathcal{A}(0,\infty),\Omega)\simeq(\mathcal{A}(-\infty,+\infty),\Omega
_{2\pi})
\end{equation}
This conformal equivalence has a generalization to the restriction of the
vacuum to chiral algebras $\mathcal{A}(a,b)$ localized in arbitrary intervals;
in this case the temperature changes with the interval.

One expects the energy and entropy of the right hand side to have the usual
(one-dimensional) volume proportionality i.e. $s=ls_{2\pi}$ where $l$
corresponds to the standard volume factor and $s_{2\pi}$ to the volume
density. The unitary equivalence map intertwines the the translation of the
heat bath theory on the right hand side with the dilation on the left hand
side. In particular it transforms the length $l$ into $\varepsilon=e^{-l}$ so
that the area density in question behaves as\footnote{A more refined analysis
reveals that the attenuation length $\varepsilon$ is really a short hand
notation for a unharmonic conformally invariant ratio.}
\begin{equation}
s_{area}=\left\vert \ln\varepsilon\right\vert s_{2\pi}+finite,~\varepsilon
\rightarrow0
\end{equation}
The remaining problem consists in verifying the $l-$proportionality and
computing the coefficient $s_{2\pi}$ in its dependence on the data of the
chiral model. This is achieved by approximating the divergent entropy of the
heat bath system by the high temperature limit of a rotational system where
the temperature is interpreted as a radius whose size is related to $l$). This
"\textit{relativistic box quantization}", which constitutes the second step,
holds also in higher dimensional conformal QFTs \cite{To}. The last crucial
step consists in using the \textit{temperature duality} which holds for the
rotational partition function of $\hat{L}_{0}=L_{0}-\frac{c}{24}.$ In this way
one verifies the $l-$proportionality and finds $s_{2\pi}=\frac{c}{12}.$ The
three steps have been described in more detail in \cite{CQG1}\cite{CQG2}.

A more refined formulation of the split process in which the localization
entropy of a chiral interval (a,b) is approximated from the inside by $(c,d)$
relates $\varepsilon$ with the conformally invariant cross ratio%
\begin{equation}
\varepsilon^{2}=\frac{\left(  b-a\right)  \left(  d-c\right)  }{\left(
c-a\right)  \left(  b-d\right)  }%
\end{equation}
This conformally invariant dependence instead of the volume factor could have
been introduced as a conformal refinement for for the $l$ dependence already
for the chiral heat bath entropy. $.$

Note that whereas the above "inverse Unruh effect" as well as the temperature
duality is not expected to hold beyond chiral theories, the "relativistic box"
approximation of the heat bath thermodynamic limit is well-defined in every in
every conformally invariant theory independent of spacetime dimensions.

With the insight that chiral localization entropy is equal to heat bath
entropy apart from a change in the parametrization resulting from the
conformal equivalence, the holographic localization entropy and its universal
area proportionality has been considerably demystified. The main open problem
in the application to black holes is to understand whether and how quantum
gravitation is capable to lead to a numerical value for $\varepsilon;$
according to its microscopic derivation all values of $\varepsilon$ are
consistent with the Hawking's thermal radiation. Arguments that the value can
be obtained by thermal re-interpretation of a classical area density are still
frail; the preservation of a classical value in the quantum setting would
appear totally unusual. Since the realistic derivation of the Hawking
radiation of a collapsing star cannot be done in a thermal equilibrium setting
but rather involves a stationary entropy flow, one may question the
applicability of all thermal equilibrium ideas (including the present one) to
black hole physics.

Although the relativistic box approximation is a conformal improvement of the
standard box approximation in the formulation of the thermodynamic limit, it
is desirable to have a more intrinsic formulation in which the thermodynamic
limit is approached by a sequence of genuine subsystems (Boxes are belonging
to unitary inequivalent systems which are only subsystems in a metaphorical
sense). This will be done in the next section which does not use any of the
three previous facts but is solely based on the split property. In this way
one is able work with a definition of localization entropy which in principle
is capable to describe the dependence on the attenuation cloud for finite
$\varepsilon$ and not only the leading terms (in the heat bath case the box
quantization is only trustworthy in its leading volume term).

\section{Localization entropy via the split density matrix}

The second approach to localization entropy also draws its strength from
chiral simplifications, but instead of conformally connecting the localization
thermality of a chiral system to its heat bath KMS properties via the somewhat
metaphoric "relativistic box approximation" of the previous section addresses
it makes direct use of the split property which identifies the approximating
algebra as a bona fide subalgebra of the same mathematical description.

In the algebraic setting a QFT is fixed in terms of a space-time indexed net
of operator algebras. In the context of a chiral theory this means the net of
operator algebras indexed by proper intervals $I$ on a circle $S^{1}%
\simeq\mathbb{\dot{R}}$ where we will use the $\mathbb{\dot{R}}$ setting of
the one-point compactified line. We pick 4 points on the line $b_{1}%
<a_{1}<a_{2}<b_{2}$ and consider the algebras $A(I_{a})\subset A(I_{b})$ where
$I_{a}=(a_{1},a_{2}),$ $I_{b}=(b_{1},b_{2})$ are properly included intervals.
Under rather mild assumptions about phase-space degree of freedoms which are
certainly valid in chiral models with a finite partition function
$Z=tre^{-\tau\hat{L}_{0}}$ the split property (as studied in the second
section) is valid and leads to the following tensor factorization%

\begin{align}
&  \mathcal{A}(I_{a})\vee\mathcal{A}(I_{b}^{\prime})\simeq\mathcal{A}%
(I_{a})\otimes\mathcal{A}(I_{b}^{\prime})\\
&  B(H)=\mathcal{N}\otimes\mathcal{N}^{^{\prime}},~\mathcal{A}(I_{b}%
)\subset\mathcal{N}\subset\mathcal{A}(I_{b})\nonumber\\
&  V(\mathcal{N})\mathcal{A}(I_{a})\vee\mathcal{A}(I_{b}^{\prime
})V(\mathcal{N})^{\ast}=\mathcal{A}(I_{a})\otimes\mathcal{A}(I_{b}^{\prime
})\nonumber
\end{align}
Here $I_{b}^{\prime}$ denotes the complement of $I_{b}$ and we used Haag
duality $A(I_{b})^{\prime}=A(I_{b}^{\prime}).$ To every concrete split i.e.
the existence of an intermediate quantum mechanical type I factor between two
monads $\mathcal{A}(I_{b})\subset\mathcal{N}\subset\mathcal{A}(I_{b})$ there
exists a unique (by suitable normalization) implementer $V(\mathcal{N})$ of
the split isomorphism.

The many different splittings correspond vaguely to classical boundary
conditions, but as a result of the increase of possibilities caused by the
finite thickness $a_{1}-b_{1}$ and $b_{2}-a_{2}$ of the two boundary between
$I_{a}$ and $I_{b}^{\prime}$ there are vastly more possibilities than in the
classical case, although one expects (as for the heat bath systems in the
thermodynamic limit) that they share the leading $\ln\varepsilon$ behavior.

Mathematically there is one preferred split in which the two monads
$\mathcal{A}(I_{a})\subset\mathcal{A}(I_{b})$ uniquely determines a
"canonical" split. The formula for this type factor $\mathcal{N}_{c}$ which is
functorially determined by the two monads reads%
\begin{equation}
\mathcal{N}_{c}=\mathcal{A}(I_{a})\vee J\mathcal{A}(I_{a})J=\mathcal{A}%
(I_{b})\wedge J\mathcal{A}(I_{b})J
\end{equation}
i.e. it is the operator algebra generated by the monad $\mathcal{A}(I_{a})$
and its image under an antiunitary involution $J$ which comes from the modular
theory of the standard pair ($\mathcal{A}(I_{a})\vee A(I_{b}^{\prime}%
),\Omega_{vac}$). In case the inclusion is split one can show that the algebra
$\mathcal{N}_{c}$ defined by this formula is really a type I factor in terms
of which $H$ and $B(H)$ tensor-factorizes. The advantage of this canonical
choice is that it maintains the covariance under spacetime transformations, in
this case the conformal covariance. Since there are many more intermediate
type I subfactors with "fuzzy boundaries" than classical geometric boundary
conditions any comparison with classical theory has its limitation; but if one
looks for an analogy for the canonical functorial determination on may think
perhaps of free boundary condition.

We are interested in the density matrix $\rho$ which is obtained by the
restriction of the vacuum state to $\mathcal{N}_{c},$ a concept which was not
available on $\mathcal{A}(I_{a})$ since monads have no density matrix states
(and a fortiori no pure states). Note that $\rho$ represents a thermal Gibbs
state; the thermal KMS aspect is a property of any algebra which is either
(sharply) localized or contained in a localized algebra as $\mathcal{N}%
_{c}\subset\mathcal{A}(I_{b})~$and KMS states on type I algebras are Gibbs
states. The Hamiltonian is a operator in the factor space and can be read off
from $\rho$ i.e. it is an operator whose localization is inside $I_{b}.$

It is this step which replaces the somewhat artistic arguments based on
functional integrals, the rest we take from the innovative and inspiring work
of condensed matter physicists who use the field theoretic setting of
factorizing models. In spite of the intrinsicness in the definition of $\rho,$
I would presently not be able to write down an explicit formula for the
canonical $\rho(b_{1},a_{1,}a_{2},b_{2})\in\mathcal{A}(I_{b})$ even though it
is conformally covariant according to its functorial construction. But if one
wants to extract the entropy from that thermal density matrix one may first
use the \textit{replica trick} to compute $tr\rho^{n}$ for n=1,2,.... and from
there a representation of the entropy \cite{Ca} in terms of an differentiation
with respect to n at n=0. After legitimizing the uniqueness of the analytic
continuation in n by checking the prerequisites of Carlson's theorem one
obtains
\begin{equation}
s=-tr\rho\ln\rho=\frac{d}{dn}tr\rho^{n}|_{n=0}%
\end{equation}
The conformal invariance of these traces follows from the conformal covariance
of $\rho$ which in turn is a result of the functoriality of its construction
in terms of conformally covariant algebras and the conformal invariance of the
vacuum. As in the previous section this forces the traces and hence the
entropy to be a function of the cross ratio of the four end-points.

In order to avoid confusions it should be stressed that these four points are
\textit{not} to be thought of as end points of localization regions but rather
as parameters which designate a sharp localization region $I_{a}$ together
with an attenuation region for vacuum polarization given by the complement
$I_{b}~\backslash~I_{a}.$ The shape of the fuzzy attenuation cloud is
completely fixed by the canonicity of the above split procedure in terms of
the modular object associated with the canonical split of the inclusion of two
monads $(\mathcal{A}(I_{a})\subset\mathcal{A}(I_{b}),\Omega).$

This is of course much more than the method of the previous section can
deliver because the thermodynamic limit approximation by (relativistic) boxes
can only be trusted in the leading volume (here length) proportionality which
according to the previous section passes in chiral theories to the logarithm
of the in the attenuation length $\varepsilon=\frac{1}{r}$ (with r given by
the cross ratio below (\ref{cross})\footnote{In fact the conformal invariance
of the chiral entropy permits to generalize the thermodynamic limit by limits
in which the right and left hand side approach infinity with different
velocities.}) via a conformal transformation to the logarithm ). The higher
corrections from vacuum polarizations are only accounted for by the split
property and the associated canonical attenuation picture.

Such a simple correspondence between quantum heat bath- and quantum
localization- thermality is only valid in chiral theories. Whereas this is not
sufficient to relate heat bath and localization aspects in higher dimensional
QFTs, it does just that for the holographic projections.

Unfortunately the present state of mathematical technology in operator
algebras only permits to compute the leading term in the vanishing attenuation
length i. e. in praxis one presently does not obtain more than in the previous
section. But since the method is quite interesting and allows us to make
contact with recent results from condensed matter physics as in (\cite{Ca} and
references cited therein), we will present it in the sequel.

The next step in the derivation consists in the use of the \textit{replica
trick}. In the algebraic setting one starts from an n-fold tensor product of a
chiral observable algebra on the circle. The following two formulas denote
cyclic and permutation orbifold associated to the tensor product.%
\begin{align}
&  (\mathcal{A\otimes A}\otimes....\otimes\mathcal{A)}^{\mathbb{Z}_{n}}\\
&  (\mathcal{A\otimes A}\otimes....\otimes\mathcal{A)}^{P_{n}}\nonumber
\end{align}
whose construction requires the split property. It was introduced in
\cite{Lo-Fe} as an auxiliary tool to analyze problems with multi-interval
inclusions. The second line denotes the closely related permutation orbifold
whose irreducible representations are similar. \ The representation theory for
tensor products is defined with the above split map but in order to come to a
splitting situation we first apply a map which transforms an interval
$I\subset S$

As usual the Riemann surface associated with $\sqrt[n]{z}$ is the $n$-fold
ramified cover of $\mathbb{C~}\backslash~\left\{  0\right\}  .$ We may use
this as for the definition in order to map its $n$-fold ramified covering of
the Moebius group into the following subgroup of Diff(S$^{1}$) formally
written as%
\begin{equation}
z\rightarrow\sqrt[n]{\frac{\alpha z^{n}+\beta}{\bar{\beta}z^{n}+\bar{\alpha}}}
\label{diff}%
\end{equation}
The representations of the $Z(n)$ orbifold are constructed from the n right
inverses of $f(z)=z^{n}$ which are injective maps $g_{0},g_{1},..g_{n-1}$ of
$\mathbb{R\rightarrow}S^{1}$ which remain comtinuous at $\pm\infty.$ On each
interval $I\subset\mathbb{R}$ these maps are unitarily implemented and the
resulting net $\Phi_{g_{i},I}(\mathcal{A})$ can be used to define a
representation of the tensor product algebra $\mathcal{A(}I\mathcal{)}$
$\otimes...\otimes\mathcal{A(}I\mathcal{)}$ as%
\begin{equation}
\pi_{f,I}\equiv\chi_{I}\cdot(\Phi_{g_{0},I}\otimes...\Phi_{g_{n-1},I})
\end{equation}
where $\chi_{I}$ is the natural isomorphism from $\mathcal{A(}I\mathcal{)}$
$\otimes...\otimes\mathcal{A(}I\mathcal{)}$ to $\mathcal{A(}I_{0}\mathcal{)}$
$\vee...\vee\mathcal{A(}I_{n-1}\mathcal{)}$ from the canonical implementation
of the split property. The net $\pi_{f,I}$ defines a soliton of $\mathcal{A}%
_{0}$ $\otimes...\otimes\mathcal{A}_{0}$ where the subscript is a reminder
that the circle has been punctured at $\infty.$

It turns out that the restriction to the cyclic orbifold i.e. the restriction%
\begin{equation}
\tau_{f}\equiv\pi_{f}|_{\left(  \mathcal{A}\otimes...\otimes\mathcal{A}%
\right)  ^{Z_{n}}}%
\end{equation}
has an extension to the full circle i.e. is a conformal field theory
(indicated by omitting the subscript). It is quite common that a soliton
representation passes to an ordinary representation. In the case at hand the
irreducible soliton representation decomposes into a direct sum of n
diffeomorphism covariant representations $\tau_{f}$ $^{(0)},...,\tau
_{f}^{(n-1)}$ whose statistical dimension and scale dimensions (of their
generating fields) were determined in \cite{Lo-Fe}. The anomalous spin
spectrum can be red off directly from the embedding of the n-fold covering of
the Moebius group into the Diff($S^{1}$) (\ref{diff}). For the following we
only need the lowest scale dimension is%

\begin{equation}
d_{n}=\frac{n^{2}-1}{12n}c \label{d}%
\end{equation}
where c is the Virasoro constant.

The purpose of the orbifold representation in the present context is to
identify the n$^{th}$ power of the $I_{b}$-localized density matrix $\rho^{n}$
with an operator in the $\tau_{f}$ representation and to extract information
of the singular behavior for coalescent points when $I_{b}\rightarrow I_{a}$
by using the fact that the \ singularities of the branch points of this
singular limit are determined by the lowest dimensional "twist" fields of the
$Z(n)$ orbifold with dimension (\ref{d}). This is precisely what Cardy et al.
\cite{Ca} arrive after (metaphoric) use of functional arguments in order to
implement the replica trick.

The remaining steps are identical to theirs. For the cross ratio $r$ we
choose
\begin{equation}
r=\frac{\left(  a_{2}-a_{1}\right)  (b_{2}-b_{1})}{\left(  a_{1}-b_{1}\right)
(b_{2}-a_{2})} \label{cross}%
\end{equation}
which becomes singular in the limit $I_{b}\rightarrow I_{a}.$ Unfortunately
$tr\rho^{n}$ is a function of $r$ which, although uniquely fixed in terms of
$I_{a}\subset I_{b},~$is presently out of reach of our computational
abilities. Its singular behavior leads to the formula
\begin{align}
tr\rho^{n}  &  =r^{\frac{n^{2}-1}{24n}c}F_{n}(r)\\
F_{1}(r)  &  =1\nonumber
\end{align}
where the singular branch point behavior has been split off. Assuming
finiteness of the derivative $\frac{d}{dn}F_{n}(r)|_{n=1}\equiv G(r)$ at
$r\rightarrow\infty$ one obtains the limiting formula of the condensed matter
literature%
\begin{align}
&  -tr\rho\ln\rho=\frac{c}{12}\ln\frac{\left(  a_{2}-a_{1}\right)  ^{2}%
}{\varepsilon^{2}}\text{ }\\
\varepsilon &  =a_{1}-b_{1}=b_{2}-a_{2}\rightarrow0\nonumber
\end{align}

This is not the first time I have used the split property for the calculation
of localization entropy. In \cite{IJMP} I used a formula for the unitary
implementation of the splitting transformation which is limited to free
fields. The resulting leading logarithmic dependence in the attenuation depth
of the vacuum polarization led me to expect that this behavior is generic. In
order to show this I looked for other ways and found the relation with the
thermodynamic limit formula of the previous section. But it was only after I
recently became aware of the work in condensed matter physics that I succeeded
to complete my old program of computing at least the leading behavior of the
canonically defined localization entropy.

In order to avoid misunderstanding, it is not our intention to compete with
the beautiful results obtained about localization entropy in condensed matter
physics \cite{Ca}; my main point is a methodological. Functional integrals,
even in cases where they exist and are backed up by measure theory, as for
superrenormalible QFT (finite wave function renormalization), are unsuitable
for the description of localized subtheories as needed to define localitation
entropy or localization energy. They are in fact blind against the thermal
manifestations resulting from the local monad structure of localized algebras
as compared to the quantum mechanical structure of the global algebra. Monads
only occur in QFT and not in QM and functional integrals have the same
appearance in QFT and QM.

In \cite{Ca} the functional integral representation is only used in a
metaphoric way in order to implement the replica idea. All the calculations
are done in the bootstrap-formfactor setting. Indeed the setting of functional
integrals is the most marvelous metaphoric instrument of QFT. For the purpose
for which it is used by Cardy et al. it is particularily suitable, and the
fact that factorizing models are outside the range of validity of functional
integral representations will not leave a pragmatically inclined quantum field
theorist sleepless as long as his consistency checks work.

But even the staunchest pragmatist cannot fail to perceive the deep irony
which lies in the fact that in those cases where the functional integral is
exact, namely in QM, it is not possible to teach a normal course on QM using
only functional integration\footnote{It is an interesting intelectual exercise
and a test of one's conceptual understanding of QM to contemplate how quantum
theory would have evolved if Feynman's approach would have appeared before
Heisenberg's. The idea is not as harbrained as it appears at first sight
because the functional integral approach is conceptually much closer to the
old Bohr-Sommerfeld QM than Heisenberg's rather abstract setting. For
calculating quasiclassical approximations the Feynman approach is the most
elegant and effective starting point.}; on the other hand modern textbooks
tend to equate the definition of QFT with functional integral quantization
despite its metaphoric content. As a result there are particle physicists who
think that perturbative divergencies and their renormalization via cutoffs or
regulaters are intrinsic attributes of QFT. It is often not noticed that the
causal approach has shown already many decades ago that the principles of QFT
implemented iteratively, starting with the Wick-ordered lowest order
interaction density, lead to a finite formulation which however in certain
cases has an increasing number of free parameters (nonrenormalizability) and
as a result ceases to be useful.

Though in most cases (including the present one) one does not really have to
rely on metaphors, their use often significantly facilitate the communication
between particle physicists. Writing a specific functional integral on a
blackboard generates a strain of associations which is generally sufficient to
initiate a meaningful discussion; it is hard to think of any other compact
effective way. The metaphoric power is strongest when the setting is used as a
vehicle to discover new mathematical structures as it was first done in the
work by Atiyah and Witten in the 70/80s.

By during the last two decades the limitations of this metaphoric power having
become increasingly evident. The local covariance principle in the context of
QFT in generic curved spacetimes is not even metaphorically compatible with a
functional integral setting, and neither are QFT with braid group statistics
as chiral models. Also there are factorizing models which are metaphorically
consistent with a functional representation most of them are not; and even if
they are, as e.g. the Sine-Gordon model, the functional metaphor is of no help
in its solution.

The present state of QFT is that of an ongoing paradigmatic change where at
the end one expects to arrive at a setting which parallels the conceptual
cohesion and the mathematical precision of the operator formalism of QM.
During this transition time the functional integral setting will continue to
be the source of new ideas. There is no harm in using its suggestive power as
long as one remains aware that it is of a metaphoric nature.

There is an interesting conceptual difference which remains between my work on
localization entropy and the work by condensed matter physicists even though
both used a QFT framework. From my point of view the use of the terminology
"cutoff" in connection with localization entropy is not helpful because its
creates the wrong association; for this reason I have avoided it ever since I
started my work at the beginning of this decade \cite{IJMP} and used instead
the concept of an attenuation length $\varepsilon$. Hardly anybody would
associate the divergent volume factor which appears in the thermodynamic limit
of thermal systems with a cutoff, yet the attenuation length parameter of the
vacuum polarization cloud is nothing but a conformal transform of the length
factor $L$ which appears in the thermodynamic limit of a chiral heat bath QFT.

Cutoffs in QFT are uncontrollable changes of theories caused by cutting out
the high energy contributions in certain integrations in the hope that despite
the uncontrollable change certain numerical quantities of interests may change
only little. The notion of attenuation length for localization-caused vacuum
polarization on the other hand is a rigorous concept \textit{within} each
fixed QFT model.

\section{The conceptual-philosophical basis of a modular-based approach,
messages for QG}

A radically different approach to QFT as the present one, which substitutes
any kind of quantization parallelism to classical fields by completely
autonomous concepts should come with different conceptual-philosophical
message of what constitutes the essence of QFT. Indeed the scenario of
holography and its inversion via reconstruction wedge asks for a different
philosophical setting than that of Lagrangian \ quantization. Whereas similar
to QM the latter harmonizes with a Newtonian view of quantum matter as
something that fills spacetime, the monad structure of local operator algebras
in QFT and their intersection and generating properties require Leibniz's more
abstract view of spacetime as an ordering device such that holography is a
radical change of this ordering device.

The underline the radical aspect of this new viewpoint we refer to
\cite{interface} where it was pointed out that quantum matter
\textit{together} with its spacetime symmetries as well as all its inner
symmetries can be encoded into the \textit{position of a finite number of
monads} (i.e. copies of the unique abstract monad) \textit{in a common Hilbert
space} \cite{interface}. \ Even the kind of quantum matter (hadrons, leptons,
photons) is resolved in terms of positioning, with other words the ultimate
reality of QFT is relative positioning of a finite number of monads in a
Hilbert space.  What makes this different perspective of QFT so interesting is
that it is completely rigorous as well as conservative. It does not replace or
add any physical principle, yet it implies a strong change of paradigm. This
is the strongest indication yet that QFT is still a very young theory with
expected changes and certainly nowhere near to its closure.

In the form as applied in this paper modular theory was used to address the
localization problems and symmetries of QFT in Minkowski spacetime. It is
natural to ask whether these ideas using modular groups can be applied in the
more general context of QFT in CST. A more modest question in this direction
would be to understand whether the Diff($S^{1}$) symmetries beyond Moebius
symmetries which do not preserve the vacuum can be obtained by modular methods
(i.e. without assuming the existence of an energy-momentum tensor which for
chiral theories originating from holography is in any case not a reasonable
assumption). Preliminary studies indicate that this is the case if one relaxes
some of the modular concepts.

An important issue is how to view the generic area proportionality of
localization entropy of quantum matter on null-horizons in connection with
Bekenstein's classical area behavior in Einstein-Hilbert like classical field
theories. The standard argument consists in using Bekenstein's quantum
re-interpretation as a key to learn something about the elusive QG. Whatever
one wants to use it for, one can certainly not claim that the entropy area law
is direct evidence of manifestation of QG. The thermal aspects of Hawking
radiation as well as the area proportionality of entropy are perfectly
describable in the setting of QFT in CST; no appeal to a still elusive QG is necessary.

The formation of a black hole through a collapsing star, as envisaged by
Hawking \cite{Haw} and described in more detail within an algebraic QFT
setting by Haag and Fredenhagen \cite{Ha-Fre}, is outside the static
equilibrium thermodynamic setting. For such stationary nonequilibrium states
the recent notion of entropy flux in the operator algebra setting \cite{Ja-Pi}
may be more appropriate.

\textbf{Acknowledgement}: I thank Henning Rehren for a helpful comment.

\end{document}